\documentclass[%
 reprint,
 amsmath,amssymb,
 aps,
prb,
superscriptaddress,
]{revtex4-1}

\usepackage{graphicx}
\usepackage{dcolumn}
\usepackage{bm}
\usepackage{hhline}
\usepackage{float}
\usepackage[utf8]{inputenc}
\usepackage[english]{babel}
\usepackage{color}

\begin{document}

\preprint{APS/123-QED}

\title{Hole-doped cobalt-based Heusler phases as prospective high-performance high-temperature thermoelectrics}
\author{Mohd~Zeeshan}
\affiliation{Indian Institute of Technology Roorkee, Department of Chemistry, Roorkee 247667, Uttarakhand, India}
\author{Jeroen van den Brink}
\affiliation{Institute for Theoretical Solid State Physics, IFW Dresden, Helmholtzstrasse 20, 01069 Dresden, Germany}
\author{Hem C. Kandpal}
\affiliation{Indian Institute of Technology Roorkee, Department of Chemistry, Roorkee 247667, Uttarakhand, India}

\date{\today}

\begin{abstract}
Materials design based on first-principles electronic calculations has proven a fruitful strategy to identify new thermoelectric materials with a favorable figure of merit. Recent electronic structure calculations predict that in cobalt-based half-Heusler systems a power factor higher than in CoTiSb can be achieved upon \textit{p}-type doping of CoVSn, CoNbSn, CoTaSn, CoMoIn, and CoWIn. Here, using a first-principles approach and semi-classical Boltzmann transport theory, we investigate the electrical and thermal transport properties of these materials. The calculated thermal conductivity at room temperature of all the systems is lower than that of CoTiSb, with CoMoIn and CoWIn having an almost 3-fold lower thermal conductivity than CoTiSb. We also provide conservative estimates of the figure of merit for these systems which all turn out to be higher than in CoTiSb and to have a maximum value for CoWIn. 
\end{abstract}

\maketitle

Recently, half-Heusler (hH) alloys, \textit{XYZ}, with cubic structure (\textit{F$\bar4$3m}) have been reported to exhibit substantial thermoelectric (TE) properties comparable to conventional Bi$_2$Te$_3$ and PbTe based materials \cite{Yu2009, Joshi2011, Yan2012, Sakurada2005, Li2016, Fang2016}. The thermal stability, mechanical strength, and semiconducting behavior of 18 valence electron count (VEC) hH alloys make them promising prospects for TE applications. Despite having a high power factor (PF), S$^2\sigma$, the high thermal conductivity lowers the efficiency of hH alloys \cite {Misra2014, Poon2001, Chen2013, Huang2016, Xie2012}. In recent years, different approaches have been adopted to enhance the efficiency of hH alloys. Along with improving the TE properties of existing materials, the search for new potential hH alloys is being carried out extensively. The interplay of theory and experiment for designing new materials had been fruitful in the past. The theoretical prediction of stable systems allows the experimentalists to narrow down the window for targeting new materials. Recently, along with experimental realization, quite a large class of hH alloys have been predicted to be theoretically stable \cite{Gautier2015, Zakutayev2013}. This motivated us to investigate specifically 18 VEC hH alloys, which were not studied in depth so-far.
  
In a previous work, we screened the cobalt based 18 VEC hH systems CoVSn, CoNbSn, CoTaSn, CoCrIn, CoMoIn, and CoWIn\cite{Zeeshan2017}, using the first-principles approach. We systematically investigated the static and dynamic stability, electronic structure, and electrical transport properties of these systems. These calculations suggested that all these systems, except for CoCrIn, could be potential TE materials on \textit{p}-type doping. Assuming relaxation time, \textit{$\tau$} = 10$^{-15}$~s, a maximum power factor (PF) was calculated of 21, 23, 22, 11, and 15 $\mu$W/K$^2$~cm for CoVSn (at 0.26 \textit{p}-type doping), CoNbSn (at 0.21 \textit{p}-type doping), CoTaSn (at 0.23 \textit{p}-type doping), CoMoIn (at 0.03 \textit{p}-type doping), and CoWIn (at 0.03 \textit{p}-type doping), respectively \cite{Zeeshan2017}. To establish the thermoelectric potential of any material, it is of key importance to evaluate the efficiency of a TE material given by the dimensionless figure of merit \textit{ZT} = \textit{S$^2\sigma$T/$\kappa$} (\textit{$\kappa$} = \textit{$\kappa_e$} + \textit{$\kappa_l$}), where S is Seebeck coefficient, $\sigma$ is electrical conductivity, and $\kappa$ is thermal conductivity comprising of electronic and lattice contributions \cite{Fang2017, Zhang2016, Hong2016, Fu2016}. 

Here we calculate the thermal conductivity, and thereby determine \textit{ZT}, of \textit{p}-type doped CoVSn, CoNbSn, CoTaSn, CoMoIn, and CoWIn combining a first-principles approach with semi-classical Boltzmann transport theory. We obtain the thermal conductivity by solving linearized Boltzmann transport equation (BTE) within the single-mode relaxation time approximation (SMA)\cite{Ziman1960} using thermal2\cite{thermal2} code as implemented in the Quantum ESPRESSO package\cite{espresso}. We employed generalized gradient approximation (GGA) given by Perdew-Burke-Ernzerhof (PBE)\cite{Perdew1996} for exchange-correlation functional. Further, we have chosen Troullier-Martins norm-conserving pseudopotentials from the Quantum ESPRESSO webpage\cite{QEweb}. An energy cutoff of 100~eV is used for the plane-wave basis set and Brillouin zone integration is performed on a Monkhorst-Pack 20$\times$20$\times$20.

It was established previously that the maximum PF for CoTiSb is obtained upon \textit{p}-type doping. Our calculations of the PF/\textit{$\tau$} as a function of \textit{p}-type doping for this system at 300, 700, and 900~K is shown in Fig. 1(a). The effect of doping on Seebeck and electrical conductivity reflects that the maximum PF is obtained when the Fermi level is near the band edge, which is indeed the case. The peak value of calculated PF/\textit{$\tau$}, at 300~K is obtained for 0.12 \textit{p}-type doping, at 700~K for 0.20 \textit{p}-type doping, and at 900~K for 0.22 \textit{p}-type doping per unit cell. This is in good agreement with previously calculated and reported values\cite{Yang2008, Wu2007}. As the predicted doping levels are in agreement with previously reported values and the optimal doping levels should be quite achievable experimentally, we proceed next to see determine the PF, thermal conductivity and \textit{ZT} as a function of temperature. Since, we obtained the maximum PF at 900~K for 0.22 \textit{p}-type doping, we focus in the following on 0.22 \textit{p}-type doped CoTiSb.  

\begin{figure}
\centering
 \includegraphics[scale=0.40]{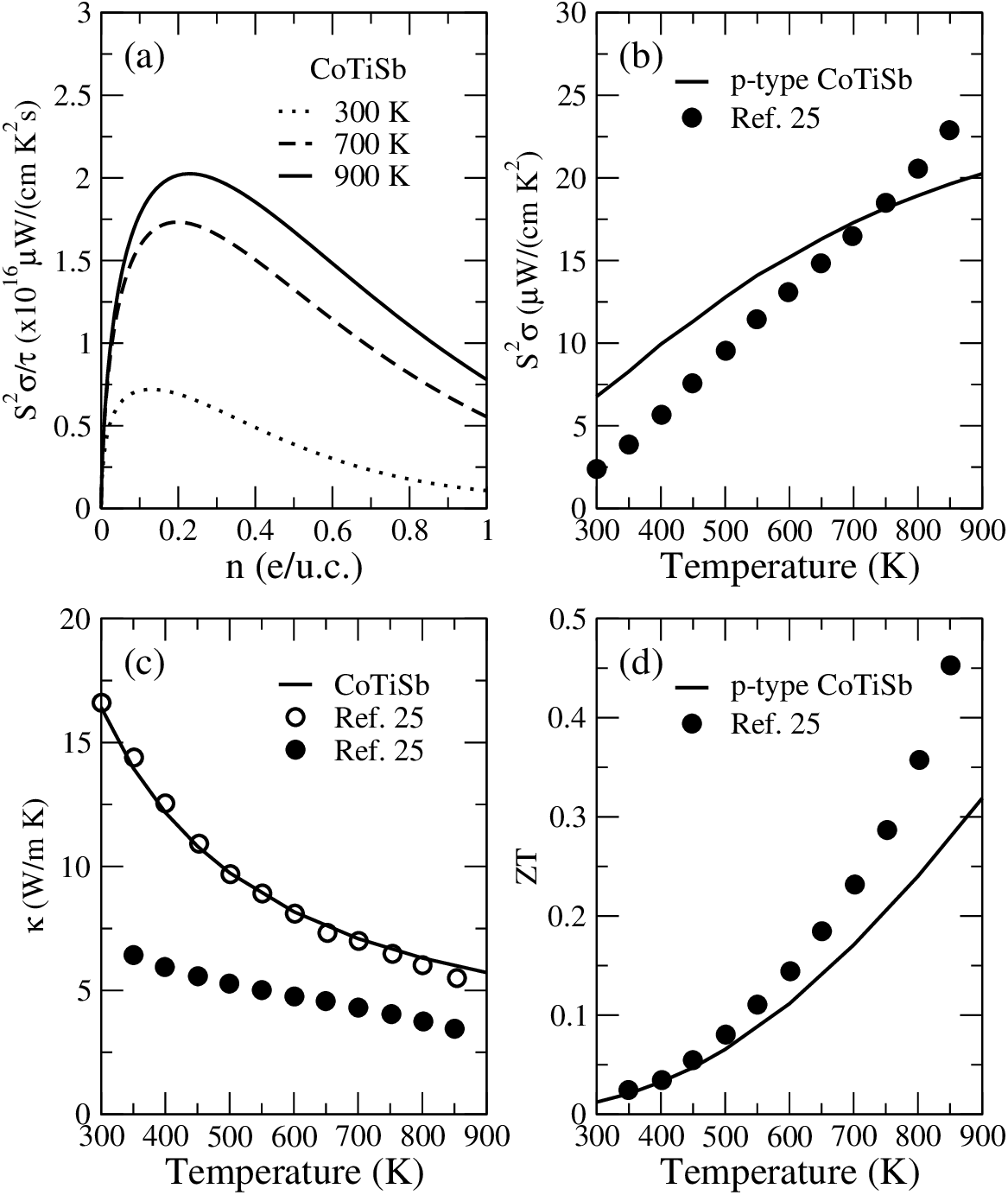}
 \caption{(a) Trend of power factor, with respect to relaxation time, on \textit{p}-type doping CoTiSb at 900~K. (b) Power factor as a function of temperature for calculated 0.22 \textit{p}-type doped CoTiSb and reported 0.15 Fe-doped CoTiSb, respectively. (c) Thermal conductivity as a function of temperature for calculated CoTiSb, reported CoTiSb, and reported 0.15 Fe-doped CoTiSb, respectively. (d) \textit{ZT} as a function of temperature for calculated 0.22 \textit{p}-type doped CoTiSb and reported 0.15 Fe-doped CoTiSb, respectively. Reported parent CoTiSb and 0.15 Fe-doped CoTiSb are represented by open and filled circles, respectively.}
\end{figure}

Fig. 1(b) shows the PF as a function of temperature for calculated 0.22 \textit{p}-type doped CoTiSb and reported 0.15 Fe-doped CoTiSb \textit{i.e.}~Co$_{0.85}$Fe$_{0.15}$TiSb\cite{Wu2007}. Assuming \textit{$\tau$} = 10$^{-15}$~s, the trend of calculated PF for 0.22 \textit{p}-type doped CoTiSb and reported PF for 0.15 Fe-doped CoTiSb is quite similar. Both calculated and reported PF values are continuously increasing with temperature. The slight deviation in magnitude of PF is because our calculated values are for the pristine system.
After obtaining a similar trend for our calculated and reported PF for the \textit{p}-type doped system, we calculate the thermal conductivity for CoTiSb next. Thermal conductivity comprises of two parts, electronic $\kappa_e$, and lattice $\kappa_l$, respectively. $\kappa_e$ is calculated by BoltzTraP code\cite{Madsen2006} whereas $\kappa_l$ is calculated by Quantum ESPRESSO. Then, we summed up both parts to obtain the total thermal conductivity, $\kappa$. The magnitude of $\kappa_e$ is too small in comparison to $\kappa_l$ that the inclusion of relaxation time becomes immaterial (Fig. 3). This is, in fact, the observed behavior of $\kappa$ in previous works \cite{Wu2009, Kimura2008, He2014}. Thus, the $\kappa_l$ plays a major contribution to total thermal conductivity.

The trend of $\kappa$ as a function of temperature for calculated CoTiSb, reported CoTiSb, and reported 0.15 Fe-doped CoTiSb is shown in Fig. 1(c). The value of calculated $\kappa$ is approximated to 10-folds lower to match the experimental results. Both calculated and reported $\kappa$, as expected, are decreasing with temperature. The calculated and reported $\kappa$ for parent CoTiSb are in good agreement. The $\kappa$ of reported 0.15 Fe-doped CoTiSb is almost 3-folds lower than that of parent CoTiSb at 300~K. This shows the importance of doping effect for a TE material. We also compared calculated $\kappa$ for parent CoNbSn with reported CoNbSn and found a good agreement in the range of 600~K to 900~K\cite{He2016}. However, the difference becomes smaller at higher temperatures. After estimating the values of PF and $\kappa$ for 0.22 \textit{p}-type doped CoTiSb, as a next step, we calculate the efficiency \textit{i.e.} \textit{ZT} for the same.

Fig. 1(d) shows the \textit{ZT} as a function of temperature for our calculated 0.22 \textit{p}-type doped CoTiSb and reported 0.15 Fe-doped CoTiSb. The \textit{ZT} values are increasing with temperature owing to the reduction in $\kappa$ at higher temperatures. A nice agreement can be seen in the range of 300~K to 600~K. Beyond that, experimental values dominate our calculated values. This is quite expected since to calculate \textit{ZT} (=\textit{S$^2\sigma$T/$\kappa$}) of \textit{p}-type doped CoTiSb, we incorporated the $\kappa$ values of parent CoTiSb. As we know, $\kappa$ decreases on doping; we expect our calculated \textit{ZT} values to lag behind the experimental values. The maximum \textit{ZT} values are 0.31 and 0.45 for calculated 0.22 \textit{p}-typed doped CoTiSb and reported 0.15 Fe-doped CoTiSb, respectively. We can say that our calculated \textit{ZT} for 0.22 \textit{p}-typed doped CoTiSb is slightly underestimated in comparison to reported 0.15 Fe-doped CoTiSb.
  
\begin{figure}
\centering
 \includegraphics[scale=0.35]{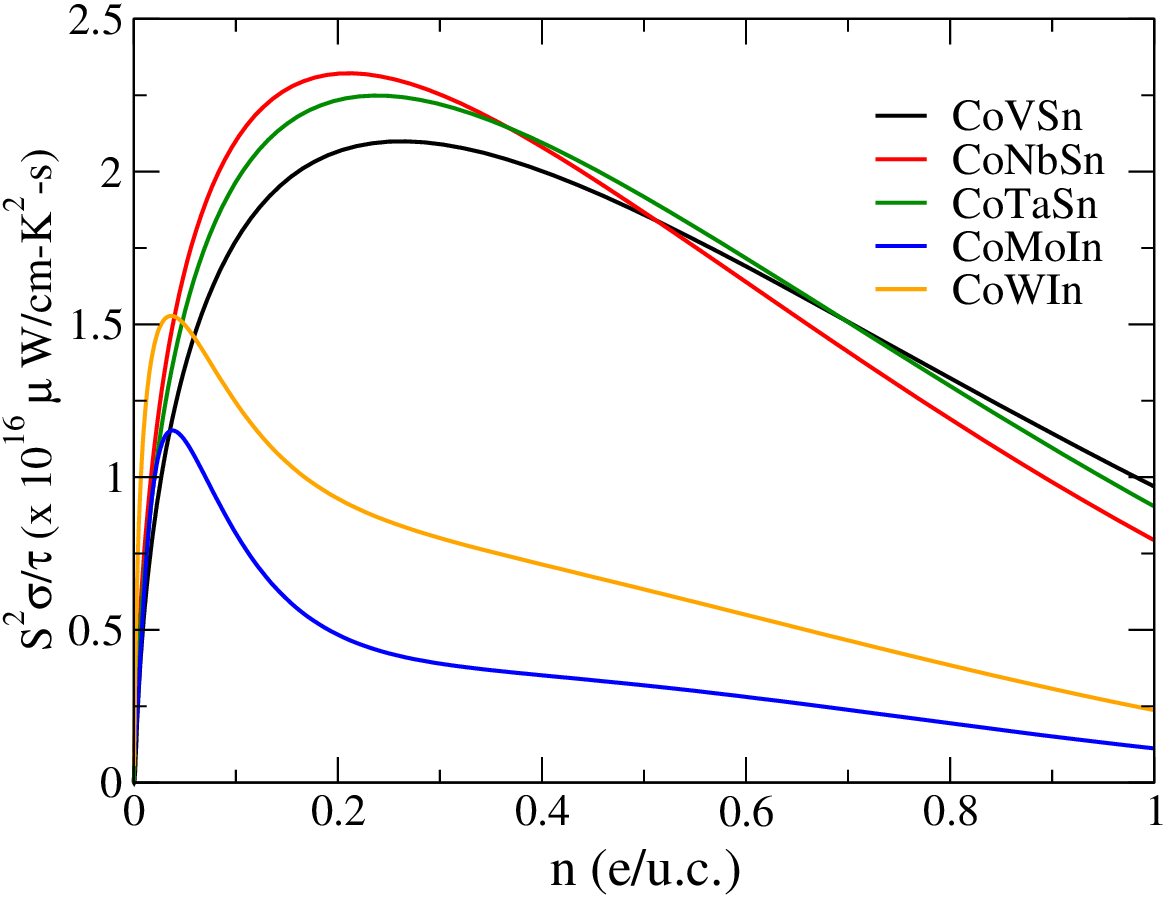}
 \caption{(Color online) The trend of power factor, with respect to relaxation time, on \textit{p}-type doping CoVSn, CoNbSn, CoTaSn, CoMoIn, and CoWIn, respectively at 900~K.}
\end{figure}

The slight deviations in our calculated values for 0.22 \textit{p}-type doped CoTiSb and reported 0.15 Fe-doped CoTiSb are expected. This is because we have calculated electrical transport properties for \textit{p}-type doped systems, in general, rather than considering a particular atom for doping. For instance, one could tune the properties by doping Co by Fe, Ru, Mn or V, or by doping Sb by Sn, Ge, or In, and properties may vary depending on the choice of atoms. Overall, we obtain a nice agreement between our calculated TE parameters and reported values for \textit{p}-type doped CoTiSb. And we proceed next to our chosen systems to analyze their TE properties.

The trend of PF/\textit{$\tau$} as a function of \textit{p}-type doping for CoVSn, CoNbSn, CoTaSn, CoMoIn, and CoWIn, respectively at 900~K is shown in Fig. 2. For all the systems, maximum PF is obtained on \textit{p}-type doping and all optimal doping levels are small and could be achieved experimentally. Table~\ref{tab} lists the optimal doping levels at which maximum PF is obtained, along with maximum PF and ZT for all the systems at 900~K. As can be seen from Table~\ref{tab}, the maximum PF is obtained for CoNbSn, followed by CoTasn, and then CoVSn. Here, we would like to propose that \textit{p}-type doped CoVSn, CoNbSn, and CoTaSn have higher PF than the well-known CoTiSb system. Recently, the \textit{p}-type doped CoVSn is predicted to have high thermoelectric performance\cite{Shi2017}. Also, CoNbSn is predicted to have higher power factor on \textit{p}-type doping\cite{Yang2008}. However, the PF for CoMoIn and CoWIn are low as compared to other \textit{p}-type doped systems. Nevertheless, we expect CoMoIn and CoWIn to possess lowest thermal conductivity on account of direct band gap and phonon scattering from heavy atoms of 4\textit{d} and 5\textit{d}-series.
  
\begin{figure}
\centering
 \includegraphics[scale=0.40]{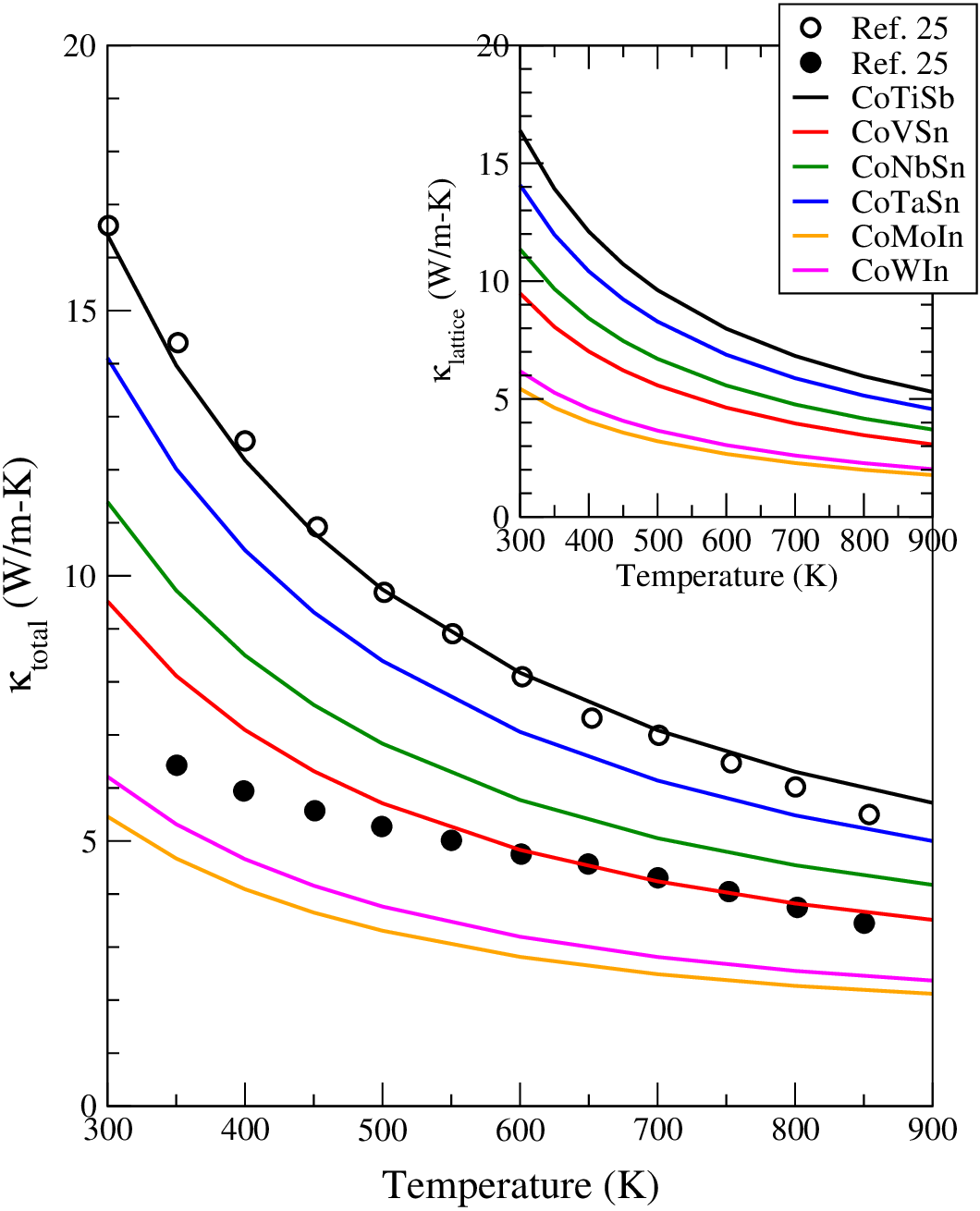}
 \caption{(Color online) The trend of thermal conductivity as a function of temperature. The inset of figure indicates the behavior of lattice thermal conductivity with temperature. Reported parent CoTiSb and 0.15 Fe-doped CoTiSb are represented by open and filled circles, respectively.}
\end{figure}

The behavior of $\kappa$ as a function of temperature for all the systems is shown in Fig. 3. The inset of Fig. 3 indicates the behavior of $\kappa_l$ with temperature. It is evident that $\kappa_l$ plays a major part in total thermal conductivity. As discussed earlier, we have approximated $\kappa$ to 10-folds lower for all the systems. The approximation seems valid since the trend for reported and calculated values for CoTiSb are in quite good agreement. As expected, $\kappa$ decreases rapidly with temperature for all the systems. The maximum $\kappa$ is obtained for CoTiSb whereas minimum $\kappa$ is obtained for CoMoIn and CoWIn. The most important observation, from Fig. 3, is that $\kappa$ for CoMoIn and CoWIn is almost 3-folds lower than that of CoTiSb at 300~K. This will significantly affect the \textit{ZT} values for CoMoIn and CoWIn, despite having a low PF as compared to other systems. 
 
Now we can determine the \textit{ZT} (=\textit{S$^2\sigma$T/$\kappa$}) values for \textit{p}-type doped systems by using the calculated $\kappa$ of parent compounds into the expression for \textit{ZT}. Since we know that $\kappa$ of the parent compound is reduced significantly on doping, we stress that our approach actually {\it underestimates} the \textit{ZT} values for all the systems. Evaluation of an exact $\kappa$ for doped systems is a subject of future work. Assuming \textit{$\tau$} = 10$^{-15}$~s, the figure of merit  \textit{ZT} for all the systems at optimal \textit{p}-type doping levels, at 900~K is shown in Table~\ref{tab}. We observe that all the proposed systems have higher \textit{ZT} than CoTiSb on \textit{p}-type doping. Despite having low PF, the maximum \textit{ZT} is obtained for CoWIn on account of lower $\kappa$. 

\begin{table}
\centering
\begin{tabular}{l|c|c|c}
System & Doping          & S$^2\sigma$          & $ZT$   \\ 
       & (e/u.c.)        & ($\mu$~W/cm~K$^2$)   &      \\ \hline
CoTiSb & 0.22            & 20.2           & 0.31 \\ \hline
CoVSn  & 0.26            & 21             & 0.53 \\ 
CoNbSn & 0.21            & 23.3           & 0.5  \\ 
CoTaSn & 0.23            & 22.4           & 0.4  \\ \hline
CoMoIn & 0.03            & 11.6           & 0.49 \\ 
CoWIn  & 0.03            & 15.2           & 0.57 \\
\end{tabular}
\caption{The maximum power factor (S$^2\sigma$) and figure of merit (\textit{ZT}) at optimum doping concentrations for \textit{p}-type doped systems at 900~K, assuming \textit{$\tau$} = 10$^{-15}$~s.}
\label{tab}
\end{table}

To summarize, all the systems show a higher power factor than well-known CoTiSb on \textit{p}-type doping. The thermal conductivity for undoped CoMoIn and CoWIn is almost 3-folds lower than that of CoTiSb at 300~K, indicative of good thermoelectric properties at room temperature. The maximum figure of merit \textit{ZT} is 0.57 for 0.03 \textit{p}-typed doped CoWIn at 900~K. The \textit{ZT} values for all the \textit{p}-type doped systems are expected to be higher than the proposed values since we have employed thermal conductivity of undoped systems. These calculations provide a clear indication for the promising thermoelectric properties of  \textit{p}-type doped CoVSn, CoNbSn, CoTaSn, CoMoIn, and CoWIn and will hopefully motivate growing these materials with optimal doping concentrations for higher performance thermoelectric applications.\\
 
{\it Acknowledgements ---}
MZ is thankful to CSIR for granting senior research fellowship. Computations were performed on HP cluster at the Computer Center for Scientific Computing (ICC), IIT Roorkee and at IFW Dresden, Germany. We thank Ulrike Nitzsche and Navneet Gupta for technical assistance. HCK gratefully acknowledge financial support from the FIG program of IIT Roorkee (Grant CMD/FIG/100596).

\end{document}